\documentclass [11pt,a4paper] {article}

\usepackage[cp1252]{inputenc}

\usepackage{amssymb}
\usepackage{amsmath}
\usepackage{amsfonts,amssymb}
\usepackage[dvips]{graphicx}
\usepackage{bbm}

\DeclareGraphicsExtensions{.pdf,.jpg,.png,.tif,.eps}

\DeclareMathAlphabet{\mathpzc}{OT1}{pzc}{m}{it}

\def\SmallColSep{\setlength{\arraycolsep}{1pt}}

\begin{document}

\title{Hardy’s paradox according to non-classical semantics}

\author{Arkady Bolotin\footnote{$Email: arkadyv@bgu.ac.il$\vspace{5pt}} \\ \textit{Ben-Gurion University of the Negev, Beersheba (Israel)}}

\maketitle

\begin{abstract}\noindent In the paper, using the language of spin-half particles, Hardy's paradox is examined within different semantics: a partial one, a many-valued one, and one defined as a set of weak values of projection operators. As it is shown in this paper, any of such non-classical semantics can resolve Hardy's paradox.\\

\noindent \textbf{Keywords:} Quantum mechanics; Hardy's paradox; Truth values; Partial semantics; Many-valued semantics; Weak values of projection operators.\\

\end{abstract}

\section{Introduction}  %{I}

\noindent In essence, Hardy's paradox can be reduced to a case of contradiction in classical logic.\\

\begin{figure}[ht!]
   \centering
   \includegraphics[scale=0.5]{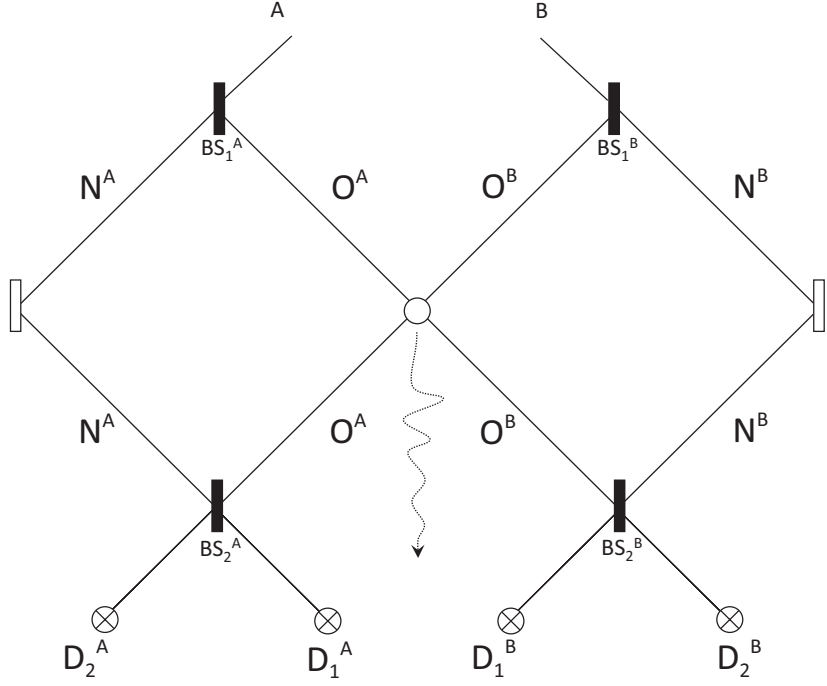}
   \caption{Diagram representing the design of Hardy’s thought experiment.\label{fig1}}
\end{figure}

\noindent Indeed, let the letters $\mathrm{A}$ and $\mathrm{B}$ denote respectively \textit{a positron} and \textit{an electron} entering their corresponding superimposed Mach-Zehnder interferometers depicted in Figure \ref{fig1} (drawn in conformity with \cite{Hardy92, Aharonov, Griffiths}). These interferometers have non-overlapping arms (denoted as $\mathrm{N}^{\mathrm{A}}$ and $\mathrm{N}^{\mathrm{B}}$) in addition to the overlapping arms (which are denoted as $\mathrm{O}^{\mathrm{A}}$ and $\mathrm{O}^{\mathrm{B}}$). Besides, each interferometer is equipped with two detectors (represented by the symbols $\mathrm{D}_1^{\mathrm{A}}$,  $\mathrm{D}_2^{\mathrm{A}}$ as well as $\mathrm{D}_1^{\mathrm{B}}$, $\mathrm{D}_2^{\mathrm{B}}$) capable of detecting the exit of the particle from the interferometer.\\

\noindent Consider at first each interferometer separately. By way of the adjustment of the arm lengths between the beam-splitters ($\mathrm{BS}_1^\mathrm{A}$ and $\mathrm{BS}_2^\mathrm{A}$ or $\mathrm{BS}_1^\mathrm{B}$ and $\mathrm{BS}_2^\mathrm{B}$) it is possible to make the particle $\mathrm{A}$ emerge only at $\mathrm{D}_2^{\mathrm{A}}$ or the particle $\mathrm{B}$ only at $\mathrm{D}_2^{\mathrm{B}}$. Then again, the detector $\mathrm{D}_1^{\mathrm{A}}$ or $\mathrm{D}_1^{\mathrm{B}}$ can be also triggered if an obstruction is present in the arm $\mathrm{O}^{\mathrm{A}}$ or $\mathrm{O}^{\mathrm{B}}$.\\

\noindent According to the design of Hardy’s thought experiment, if each interferometer is considered separately, the particle $\mathrm{A}$ can be detected only at $\mathrm{D}_2^{\mathrm{A}}$ and the particle B only at $\mathrm{D}_2^{\mathrm{B}}$. However, due to the presence of the overlapping arms in the setup of the experiment (and, for that reason, the possibility of the obstruction), the detectors $\mathrm{D}_1^{\mathrm{A}}$ and $\mathrm{D}_1^{\mathrm{B}}$ may be triggered.\\

\noindent Thus, from the clicking of $\mathrm{D}_1^{\mathrm{B}}$ one can infer that the particle $\mathrm{A}$ has gone through the overlapping arm $\mathrm{O}^{\mathrm{A}}$ obstructing the particle $\mathrm{B}$ (and because of this the particle $\mathrm{B}$ was not able to get to $\mathrm{D}_2^{\mathrm{B}}$). Similarly, the click of $\mathrm{D}_1^{\mathrm{A}}$ would mean that the particle $\mathrm{B}$ went through the overlapping arm $\mathrm{O}^{\mathrm{B}}$ obstructing the particle $\mathrm{A}$ (which, as a result, could not reach $\mathrm{D}_2^{\mathrm{A}}$).\\

\noindent Obviously, if both the particle $\mathrm{A}$ and the particle $\mathrm{B}$ went through the overlapping arms $\mathrm{O}^{\mathrm{A}}$ and $\mathrm{O}^{\mathrm{B}}$, they would annihilate. So, classically speaking, the simultaneous clicking of the detectors $\mathrm{D}_1^{\mathrm{A}}$ and $\mathrm{D}_1^{\mathrm{B}}$ is impossible.\\

\noindent Let $v$ be \textit{a valuation}, that is, a mapping from a set of propositions $\{\diamond\}$ (where the symbol $\diamond$ stands for any proposition, compound or simple) to a set of truth-values $\mathcal{V}_N = \{\mathfrak{v}\}$ having the cardinality $N$ and the range with the upper bound 1 (representing \textit{the truth}) and the lower bound 0 (representing \textit{the falsehood}). As it is customary \cite{Gottwald}, the double-bracket notation ${[\![ \,\diamond\, ]\!]}_v$ is used to denote the valuation $v$.\\

\noindent Based on the setup of the interferometers, one can derive the next equalities:\smallskip

\begin{equation} \label{1} %{Eq.1}
   {[\![ \,\mathrm{D}_1^{\mathrm{B}}\, ]\!]}_v
   =
   {[\![ \,\mathrm{O}^{\mathrm{A}}\, ]\!]}_v
   \;\;\;\;  ,
\end{equation}

\begin{equation} \label{2} %{Eq.2}
   {[\![ \,\mathrm{D}_1^{\mathrm{A}}\, ]\!]}_v
   =
   {[\![ \,\mathrm{O}^{\mathrm{B}}\, ]\!]}_v
   \;\;\;\;  .
\end{equation}
\smallskip

\noindent They mean, for example, that the proposition asserting the passage of the particle $\mathrm{A}$ through the overlapping arm $\mathrm{O}^{\mathrm{A}}$ has the same value as the proposition asserting the click of the detector $\mathrm{D}_1^{\mathrm{B}}$; in other words, these propositions are both true or both false.\\

\noindent From these equalities it immediately follows\smallskip

\begin{equation} \label{3} %{Eq.3}
   {[\![ \,\mathrm{D}_1\, ]\!]}_v
   =
   {[\![ \,\mathrm{O}\, ]\!]}_v
   \;\;\;\;  ,
\end{equation}
\smallskip

\noindent where the symbol $\mathrm{D}_1$ stands for the proposition of the simultaneous clicking of the detectors $\mathrm{D}_1^{\mathrm{B}}$ and $\mathrm{D}_1^{\mathrm{A}}$, while the symbol $\mathrm{O}$ denotes the proposition that both particles travelled through the overlapping arms $\mathrm{O}^{\mathrm{B}}$ and $\mathrm{O}^{\mathrm{A}}$ in their respective interferometers.\\

\noindent On the other hand, the situation where both particles were in the overlapping arms $\mathrm{O}^{\mathrm{B}}$ and $\mathrm{O}^{\mathrm{A}}$ cannot occur because of the annihilation process. Accordingly, in the initial state of the experiment, i.e., before the verification of the proposition $\mathrm{D}_1$, the proposition $\mathrm{O}$ is false, namely, ${[\![ \,\mathrm{O}\, ]\!]}_v=0$.\\

\noindent Let $\mathbb{P}\left[ {[\![ \,\diamond\, ]\!]}_v = 1 \right] \in [0,1]$ be the value that represents \textit{the probability that the proposition $\diamond$ is true}. Assume that the following conditions are satisfied:\smallskip

\begin{equation} \label{4} %{Eq.4}
   {[\![ \,\diamond\, ]\!]}_v
   =
   1
   \iff
   \mathbb{P}\left[ {[\![ \,\diamond\, ]\!]}_v = 1 \right]
   =
   1
   \;\;\;\;  ,
\end{equation}   

\begin{equation} \label{5} %{Eq.5}
   {[\![ \,\diamond\, ]\!]}_v
   =
   0
   \iff
   \mathbb{P}\left[ {[\![ \,\diamond\, ]\!]}_v = 1 \right]
   =
   0
   \;\;\;\;  .
\end{equation}   
\smallskip

\noindent Then, in terms of \textit{a probabilistic logic} \cite{Leblanc, Adams}, the equality (\ref{3}) would imply that the probability of the simultaneous clicking of the detectors $\mathrm{D}_1^{\mathrm{B}}$ and $\mathrm{D}_1^{\mathrm{A}}$ must be zero:\smallskip

\begin{equation} \label{6} %{Eq.6}
   {[\![ \,\mathrm{O}\, ]\!]}_v
   =
   0
   \implies
   \mathbb{P}\left[ {[\![ \,\mathrm{D}_1\, ]\!]}_v = 1 \right]
   =
   0
   \;\;\;\;  .
\end{equation}   
\smallskip

\noindent However, in accordance with quantum mechanics, the probability $\mathbb{P}\left[ {[\![ \,\mathrm{D}_1\,  ]\!]}_v = 1 \right]$ is different from zero, that is, sometimes the particles do emerge simultaneously at $\mathrm{D}_1^{\mathrm{B}}$ and $\mathrm{D}_1^{\mathrm{A}}$. This constitutes Hardy's paradox, namely,\smallskip

\begin{equation} \label{7} %{Eq.7}
   \left.
      \begin{array}{l}
         {[\![ \,\mathrm{O}\,  ]\!]}_v
         =
         0\\
         {[\![ \,\mathrm{D}_1\, ]\!]}_v
         =
         {[\![ \,\mathrm{O}\, ]\!]}_v
       \end{array}
   \right\}
   \implies
   \mathbb{P}\left[ {[\![ \,\mathrm{D}_1\,  ]\!]}_v = 1 \right]
   \neq
   0
   \;\;\;\;  .
\end{equation}
\smallskip

\noindent As one can see, the paradox arises because of the implicit assumption supposing that any proposition relating to a quantum system is \textit{either true or false}. Mathematically, it is equivalent to the statement that a logic lying behind quantum phenomena has a non-partial bivalent semantics that is defined as a set of bivaluations ${[\![ \,\diamond\, ]\!]}_v \in \mathcal{V}_2 = \{0,1\}$.\\

\noindent This suggests that the said paradox might not appear within semantics that are partial or many-valued (or both).\\

\noindent Really, consider a ``gappy'' semantics where the valuation $v$ is the function from propositions $\{\diamond\}$ into the set $\mathcal{V}_2 = \{0,1\}$ such that $v$ is not \textit{total}. In this case, even if the proposition $\mathrm{O}$ is false, some propositions, e.g., $\mathrm{D}_1$, might have no truth-values at all, which can explain the deviation of the probability $\mathbb{P}\left[ {[\![ \,\mathrm{D}_1\,  ]\!]}_v = 1 \right]$ from zero.\\

\noindent Let us demonstrate a resolution of the paradox (\ref{7}) based on the aforesaid gappy semantics as well as gapless many-valued semantics.\\

\section{Truth-value assignment of the projection operators}  %{II}

\noindent Let $\hat{P}_{\diamond}$ be the projection operator on the Hilbert space $\mathcal{H}$ associated with some quantum system such that $\hat{P}_{\diamond}$ corresponds to the proposition $\diamond$ related to this system. Assume that the valuational axiom hold\smallskip

\begin{equation} \label{8} %{Eq.8}
   v(\hat{P}_{\diamond})
   =
   {[\![ \,\diamond\, ]\!]}_v
   \;\;\;\;  ,
\end{equation}
\smallskip

\noindent where $v$ is the truth-value assignment function.\\

\noindent To find out how this function works, let us take a quantum system prepared in a pure state $|{\Psi}_{\alpha}\rangle$ lying in the column space (range) of the projection operator $\hat{P}_{\alpha}$. Since being in $\mathrm{ran}(\hat{P}_{\alpha})$ means $\hat{P}_{\alpha}|{\Psi}_{\alpha}\rangle = 1 \cdot |{\Psi}_{\alpha}\rangle$, one can assume that in the state $|{\Psi}_{\alpha}\rangle \in \mathrm{ran}(\hat{P}_{\alpha})$, the truth-value assignment function $v$ assigns the truth value 1 to the projection operator $\hat{P}_{\alpha}$ and, in this way, the proposition $\alpha$, specifically, $v(\hat{P}_{\alpha}) = {[\![ \,\alpha\, ]\!]}_v = 1$. Contrariwise, if $v(\hat{P}_{\alpha}) = {[\![ \,\alpha\, ]\!]}_v = 1$, then one can assume that the system is prepared in the state $|{\Psi}_{\alpha}\rangle \in \mathrm{ran}(\hat{P}_{\alpha})$. These two assumptions can be recorded  together as the following logical biconditional:\smallskip

\begin{equation} \label{9} %{Eq.9}
   |{\Psi}_{\alpha}\rangle \in \mathrm{ran}(\hat{P}_{\alpha})
   \;\;
   \iff
   \;\;
   v(\hat{P}_{\alpha})
   =
   {[\![ \,\alpha\, ]\!]}_v
   =
   1
   \;\;\;\;  .
\end{equation}
\smallskip

\noindent On the other hand, the vector $|{\Psi}_{\alpha}\rangle$ is in the null space of any projection operator $\hat{P}_{\beta}$ orthogonal to $\hat{P}_{\alpha}$. Since being in $\mathrm{ker}(\hat{P}_{\beta})$ means $\hat{P}_{\beta}|{\Psi}_{\alpha}\rangle = 0 \cdot |{\Psi}_{\alpha}\rangle$, one can assume then\smallskip

\begin{equation} \label{10} %{Eq.10}
   |{\Psi}_{\alpha}\rangle
   \in
   \mathrm{ker}(\hat{P}_{\beta})
   =
   \mathrm{ran}(\hat{1} - \hat{P}_{\beta})   
   \;\;
   \iff
   \;\;
   v(\hat{P}_{\beta})
   =
   {[\![ \,\beta\, ]\!]}_v
   =
   0
   \;\;\;\;  .
\end{equation}
\smallskip

\noindent Bringing the last two assumptions into a union, one can write down the following bivaluations\smallskip

\begin{equation} \label{11} %{Eq.11}
   |\Psi\rangle
   \in
   \left\{
      \begin{array}{l}
         \mathrm{ran}(\hat{P}_{\diamond})\\
         \mathrm{ran}(\neg \hat{P}_{\diamond})
      \end{array}
   \right.
   \iff
   \;
   v(\hat{P}_{\diamond})
   =
   {[\![ \,\diamond\, ]\!]}_v
   \in
   \mathcal{V}_2
   \;\;\;\;  ,
\end{equation}
\smallskip

\noindent where the operation $\neg \hat{P}_{\diamond} = \hat{1} - \hat{P}_{\diamond}$ is understood as negation of $\hat{P}_{\diamond}$.\\

\noindent Suppose by contrast that the system is prepared in the state $|{\Psi}\rangle$ that does not lie in the column or null space of the projection operator $\hat{P}_{\diamond}$, i.e., $|{\Psi}\rangle \notin \mathrm{ran}(\hat{P}_{\diamond})$ and at the same time $|{\Psi}\rangle \notin \mathrm{ran}(\neg \hat{P}_{\diamond})$. Then, under the bivaluations (\ref{11}), the truth-value function $v$ must assign neither 1 nor 0 to $\hat{P}_{\diamond}$, that is, $v(\hat{P}_{\diamond}) \neq 1$ and $v(\hat{P}_{\diamond}) \neq 0$. This means that the proposition $\diamond$ associated with $\hat{P}_{\diamond}$ cannot be bivalent under the function $v$, namely, ${[\![ \,\diamond\, ]\!]}_v \notin \mathcal{V}_2$.\\

\noindent Using \textit{a gappy yet two-valued semantics}, the failure of bivalence can be described as the truth-value gaps, explicitly,\smallskip

\begin{equation} \label{12} %{Eq.12}
   |\Psi\rangle
   \notin
   \left\{
      \begin{array}{l}
         \mathrm{ran}(\hat{P}_{\diamond})\\
         \mathrm{ran}(\neg \hat{P}_{\diamond})
      \end{array}
   \right.
   \iff
   \;
   \{v(\hat{P}_{\diamond})\}
   =
   \{{[\![ \,\diamond\, ]\!]}_v\}
   =
   \varnothing
   \;\;\;\;  .
\end{equation}
\smallskip

\noindent Observe that under the bivaluations (\ref{11}), the projection operators $\hat{1}$ and $\hat{0}$ are true and false, respectively, in any arbitrary state $|\Psi\rangle \in \mathcal{H}$, namely,

\begin{equation} \label{13} %{Eq.13}
   |\Psi\rangle
   \in
   \left\{
      \begin{array}{l}
         \mathrm{ran}(\hat{1}) = \mathcal{H}\\
         \mathrm{ran}(\neg \hat{0}) = \mathcal{H}
      \end{array}
   \right.
   \iff
   \left\{
      \begin{array}{l}
         v(\hat{1}) = 1\\
         v(\hat{0}) = 0
      \end{array}
   \right.
   \;\;\;\;  .
\end{equation}
\smallskip

\noindent Therefore, under those bivaluations, the operator $\hat{1}$ can be equated with “\textit{the super-truth}” since it can be assigned the value of the truth in \textit{all admissible states} of the quantum system. Similarly, the operator $\hat{0}$ can be equated with “\textit{the super-falsity}” because it can be assigned the value of the falsity in all admissible states of the quantum system. Accordingly, one can call gappy semantics defined as the set of the bivaluations (\ref{11}) and the truth-value gaps (\ref{12}) \textit{quantum supervaluationism} (for other details of such semantics see, for example, \cite{Varzi, Keefe} and also \cite{Bolotin17, Bolotin18}).\\

\section{Hardy's paradox described in the language of spin-half particles}  %{III}

\noindent In the language of spin-half particles, particle $\mathrm{A}$'s states $|\,\mathrm{O}^{\mathrm{A}}\rangle$ and $|\,\mathrm{N}^{\mathrm{A}}\rangle$ in the overlapping and non-overlapping arms of the Mach-Zehnder interferometer can be represented by the normalized eigenvectors of Pauli matrices corresponding to the eigenvalues $+1$ and $-1$, namely,\smallskip

\begin{equation} \label{14} %{Eq.14}
   |\,\mathrm{O}^{\mathrm{A}}\rangle
   \stackrel{\text{\tiny def}}{=}
   \!
   \left[
      \!\!
      \begin{array}{c}
         1 \\
         0
      \end{array}
      \!\!
   \right]
   ,
   \;
   |\,\mathrm{N}^{\mathrm{A}}\rangle
   \stackrel{\text{\tiny def}}{=}
   \!\
   \left[
      \!\!
      \begin{array}{c}
         0 \\
         1
      \end{array}
      \!\!
   \right]
   \;\;\;\;  .
\end{equation}
\smallskip

\noindent Obviously, kets $|\,\mathrm{O}^{\mathrm{B}}\rangle$ and $|\,\mathrm{N}^{\mathrm{B}}\rangle$ can be represented in the same way.\\

\noindent Using such a representation, the projection operators corresponding to the propositions $\mathrm{O}^{\mathrm{A}}$ and $\mathrm{N}^{\mathrm{A}}$ are defined by\smallskip

\begin{equation} \label{15} %{Eq.15}
   {\hat{P}}_{\mathrm{O}^{\mathrm{A}}}
   =
   |\,\mathrm{O}^{\mathrm{A}}\rangle \langle\mathrm{O}^{\mathrm{A}}|
   =
   \!\left[
      \begingroup\SmallColSep
      \begin{array}{r r}
         1 & 0 \\
         0 & 0
      \end{array}
      \endgroup
   \right]
   \;\;\;\;  ,
\end{equation}

\begin{equation} \label{16} %{Eq.16}
   {\hat{P}}_{\mathrm{N}^{\mathrm{A}}}
   =
   |\,\mathrm{N}^{\mathrm{A}}\rangle \langle\mathrm{N}^{\mathrm{A}}|
   =
   \!\left[
      \begingroup\SmallColSep
      \begin{array}{r r}
         0 & 0 \\
         0 & 1
      \end{array}
      \endgroup
   \right]
   \;\;\;\;  .
\end{equation}
\smallskip

\noindent Consequently, the projection operator ${\hat{P}}_{\mathrm{O}}$ relating to the proposition $\mathrm{O}$ can be expressed as\smallskip

\begin{equation} \label{17} %{Eq.17}
   {\hat{P}}_{\mathrm{O}}
   =
   {\hat{P}}_{\mathrm{O}^{\mathrm{A}}}
   \otimes
   {\hat{P}}_{\mathrm{O}^{\mathrm{B}}}
   =
   |\,\mathrm{O}^{\mathrm{A}}\rangle \langle\mathrm{O}^{\mathrm{A}}|
   \otimes
   |\,\mathrm{O}^{\mathrm{B}}\rangle \langle\mathrm{O}^{\mathrm{B}}|
   =
   \!\left[
      \begingroup\SmallColSep
      \begin{array}{r r}
         1 & 0 \\
         0 & 0
      \end{array}
      \endgroup
   \right]
   \otimes
   \!\left[
      \begingroup\SmallColSep
      \begin{array}{r r}
         1 & 0 \\
         0 & 0
      \end{array}
      \endgroup
   \right]
   =
   \!\left[
      \begingroup\SmallColSep
      \begin{array}{r r r r}
         1 & 0 & 0 & 0 \\
         0 & 0 & 0 & 0 \\
         0 & 0 & 0 & 0 \\
         0 & 0 & 0 & 0
      \end{array}
      \endgroup
   \right]
   \;\;\;\;  .
\end{equation}
\smallskip

\noindent The column and null spaces of this operator are\smallskip

\begin{equation} \label{18} %{Eq.18}
   \mathrm{ran}({\hat{P}}_{\mathrm{O}})
   =
   \left\{
   \left[
      \!\!
      \begin{array}{c}
                a \\
                0 \\
                0 \\
                0
      \end{array}
      \!\!
   \right]\!
   :\,
   a \in \mathbb{R}
   \right\}
   \;\;\;\;  ,
\end{equation}

\begin{equation} \label{19} %{Eq.19}
   \mathrm{ran}(\neg {\hat{P}}_{\mathrm{O}})
   =
   \left\{
   \left[
      \!\!
      \begin{array}{c}
                0 \\
                b \\
                c \\
                d
      \end{array}
      \!\!
   \right]\!
   :\,
   b,c,d \in \mathbb{R}
   \right\}
   \;\;\;\;  .
\end{equation}
\smallskip

\noindent According to the bivaluations (\ref{11}), for the proposition $\mathrm{O}$ to have the value of the falsity in the state $|\Psi_{\neg\mathrm{O}}\rangle$, the last-named must lie in the null space of the projection operator ${\hat{P}}_{\mathrm{O}}$:\smallskip

\begin{equation} \label{20} %{Eq.20}
   |\Psi_{\neg\mathrm{O}}\rangle
   \in
   \mathrm{ran}(\neg \hat{P}_{\mathrm{O}})
   \;\;
   \iff
   \;\;
   v(\hat{P}_{\mathrm{O}})
   =
   {[\![ \,\mathrm{O}\, ]\!]}_v
   =
   0
   \;\;\;\;  .
\end{equation}
\smallskip

\noindent Provided $b=c=d$, this state $|\Psi_{\neg\mathrm{O}}\rangle$ can be written down as\smallskip

\begin{equation} \label{21} %{Eq.21}
   |\Psi_{\neg\mathrm{O}}\rangle
   =
   \frac{1}{\sqrt{3}}
   \!
   \left[
      \!\!
      \begin{array}{c}
                0 \\
                1 \\
                1 \\
                1
      \end{array}
      \!\!
   \right]
   \!
   =
   \frac{1}{\sqrt{3}}
   \bigg(
      \!
      |\,\mathrm{O}^{\mathrm{A}}\rangle \otimes |\,\mathrm{N}^{\mathrm{B}}\rangle
      +
      |\,\mathrm{N}^{\mathrm{A}}\rangle \otimes |\,\mathrm{O}^{\mathrm{B}}\rangle
      +
      |\,\mathrm{N}^{\mathrm{A}}\rangle \otimes |\,\mathrm{N}^{\mathrm{B}}\rangle
      \!
   \bigg)
   \;\;\;\;   
\end{equation}
\smallskip

\noindent and taken as the initial state of the system (i.e., the state prior to the verification).\\

\noindent The detectors in particle $\mathrm{A}$'s interferometer verify the values of the operators ${\hat{P}}_{\mathrm{D}_2^{\mathrm{A}}}$ and ${\hat{P}}_{\mathrm{D}_1^{\mathrm{A}}}$, whose projections on the states $|\,\mathrm{O}^{\mathrm{A}}\rangle$ and $|\,\mathrm{N}^{\mathrm{A}}\rangle$ can be defined by the following superpositions\smallskip

\begin{equation} \label{22} %{Eq.22}
   |{\mathrm{D}_2^{\mathrm{A}}}\rangle
   =
   \frac{1}{\sqrt{2}}
   \bigg(
      \!
      |\,\mathrm{O}^{\mathrm{A}}\rangle
      +
      |\,\mathrm{N}^{\mathrm{A}}\rangle
      \!
   \bigg)
   =
   \frac{1}{\sqrt{2}}
   \bigg(
      \!
      \left[
         \!\!
         \begin{array}{c}
            1 \\
            0
         \end{array}
         \!\!
      \right]
       + 
      \left[
         \!\!
         \begin{array}{c}
            0 \\
            1
         \end{array}
         \!\!
      \right]
      \!
   \bigg)
   =
   \frac{1}{\sqrt{2}}
   \bigg(
      \!
      \left[
         \!\!
         \begin{array}{c}
            1 \\
            1
         \end{array}
         \!\!
      \right]
      \!
   \bigg)
   \;\;\;\;  ,
\end{equation}

\begin{equation} \label{23} %{Eq.23}
   |{\mathrm{D}_1^{\mathrm{A}}}\rangle
   =
   \frac{1}{\sqrt{2}}
   \bigg(
      \!
      |\,\mathrm{O}^{\mathrm{A}}\rangle
      -
      |\,\mathrm{N}^{\mathrm{A}}\rangle
      \!
   \bigg)
   =
   \frac{1}{\sqrt{2}}
   \bigg(
      \!
      \left[
         \!\!
         \begin{array}{c}
            1 \\
            0
         \end{array}
         \!\!
      \right]
       - 
      \left[
         \!\!
         \begin{array}{c}
            0 \\
            1
         \end{array}
         \!\!
      \right]
      \!
   \bigg)
   =
   \frac{1}{\sqrt{2}}
   \bigg(
      \!
      \left[
         \!\!
         \begin{array}{r}
            1 \\
           -1
         \end{array}
         \!\!
      \right]
      \!
   \bigg)
   \;\;\;\;  .
\end{equation}
\smallskip

\noindent These superpositions (and equivalent ones for particle $\mathrm{B}$'s states $|\,\mathrm{O}^{\mathrm{B}}\rangle$ and $|\,\mathrm{N}^{\mathrm{B}}\rangle$) allow one to identify the projection operators ${\hat{P}}_{\mathrm{D}_2^{\mathrm{A}}}$ and ${\hat{P}}_{\mathrm{D}_1^{\mathrm{A}}}$ (along with ${\hat{P}}_{\mathrm{D}_2^{\mathrm{B}}}$ and ${\hat{P}}_{\mathrm{D}_1^{\mathrm{B}}}$)\smallskip

\begin{equation} \label{24} %{Eq.24}
   {\hat{P}}_{\mathrm{D}_2^{\mathrm{A}}}
   =
   |{\mathrm{D}_2^{\mathrm{A}}}\rangle \langle{\mathrm{D}_2^{\mathrm{A}}}|
   =
   \frac{1}{2}
   \!\left[
      \begingroup\SmallColSep
      \begin{array}{r r}
         1 & 1 \\
         1 & 1
      \end{array}
      \endgroup
   \right]
   \;\;\;\;  ,
\end{equation}

\begin{equation} \label{25} %{Eq.25}
   {\hat{P}}_{\mathrm{D}_1^{\mathrm{A}}}
   =
   |\mathrm{D}_1^{\mathrm{A}}\rangle \langle\mathrm{D}_1^{\mathrm{A}}|
   =
   \frac{1}{2}
   \!\left[
      \begingroup\SmallColSep
      \begin{array}{r r}
         1 & -1 \\
        -1 &  1
      \end{array}
      \endgroup
   \right]
   \;\;\;\;  ,
\end{equation}
\smallskip

\noindent and, then, the projection operators ${\hat{P}}_{\mathrm{D}_2}$ and ${\hat{P}}_{\mathrm{D}_1}$\smallskip

\begin{equation} \label{26} %{Eq.26}
   {\hat{P}}_{\mathrm{D}_2}
   =
   {\hat{P}}_{\mathrm{D}_2^{\mathrm{A}}} \otimes {\hat{P}}_{\mathrm{D}_2^{\mathrm{B}}}
   =
   \frac{1}{4}
   \!\left[
      \begingroup\SmallColSep
      \begin{array}{r r r r}
         1 & 1 & 1 & 1 \\
         1 & 1 & 1 & 1 \\
         1 & 1 & 1 & 1 \\
         1 & 1 & 1 & 1
      \end{array}
      \endgroup
   \right]
   \;\;\;\;  ,
\end{equation}

\begin{equation} \label{27} %{Eq.27}
   {\hat{P}}_{\mathrm{D}_1}
   =
   {\hat{P}}_{\mathrm{D}_1^{\mathrm{A}}} \otimes {\hat{P}}_{\mathrm{D}_1^{\mathrm{B}}}
   =
   \frac{1}{4}
   \!\left[
      \begingroup\SmallColSep
      \begin{array}{r r r r}
          1 & -1 & -1 &  1 \\
         -1 &  1 &  1 & -1 \\
         -1 &  1 &  1 & -1 \\
          1 & -1 & -1 &  1
      \end{array}
      \endgroup
   \right]
   \;\;\;\;  ,
\end{equation}
\smallskip

\noindent whose column and null spaces are\smallskip

\begin{equation} \label{28} %{Eq.28}
   \mathrm{ran}({\hat{P}}_{\mathrm{D}_2})
   =
   \left\{
   \left[
      \!\!
      \begin{array}{r}
                a \\
                a \\
                a \\
                a
      \end{array}
      \!\!
   \right]\!
   :\,
   a \in \mathbb{R}
   \right\}
   ,
   \;
   \mathrm{ran}(\neg {\hat{P}}_{\mathrm{D}_2})
   =
   \left\{
   \left[
      \!\!
      \begin{array}{r}
                -b-c-d \\
                        b \\
                        c \\
                       d
      \end{array}
      \!\!
   \right]\!
   :\,
   b,c,d \in \mathbb{R}
   \right\}
   \;\;\;\;  ,
\end{equation}

\begin{equation} \label{29} %{Eq.29}
   \mathrm{ran}({\hat{P}}_{\mathrm{D}_1})
   =
   \left\{
   \left[
      \!\!
      \begin{array}{r}
                 a \\
                -a \\
                -a \\
                 a
      \end{array}
      \!\!
   \right]\!
   :\,
   a \in \mathbb{R}
   \right\}
   ,
   \;
   \mathrm{ran}(\neg {\hat{P}}_{\mathrm{D}_1})
   =
   \left\{
   \left[
      \!\!
      \begin{array}{r}
                b+c-d \\
                       b \\
                       c \\
                       d
      \end{array}
      \!\!
   \right]\!
   :\,
   b,c,d \in \mathbb{R}
   \right\}
   \;\;\;\;  .
\end{equation}
\smallskip

\noindent As follows, the null space of the projection operator ${\hat{P}}_{\mathrm{O}}$ cannot be a subset or superset of the column space or the null space of the projection operator ${\hat{P}}_{\mathrm{D}_1}$, namely,\smallskip

\begin{equation} \label{30} %{Eq.30}
   \left\{
   \left[
      \!\!
      \begin{array}{c}
                0 \\
                b \\
                c \\
                d
      \end{array}
      \!\!
   \right]\!
   :\,
   b,c,d \in \mathbb{R}
   \right\}
   \!\!
   \begin{array}{c}
             \not\subset \\
             \not\supset
   \end{array}
   \!\!
   \left\{
      \!\!
      \begin{array}{c}
               \left\{
                  \left[
                  \!\!
                  \begin{array}{r}
                          a \\
                         -a \\
                         -a \\
                          a
                  \end{array}
                  \!\!
                  \right]\!
                  :\,
                  a \in \mathbb{R}
               \right\}
               \\
               \left\{
               \left[
               \!\!
               \begin{array}{r}
                        b+c-d \\
                               b \\
                               c \\
                               d
               \end{array}
               \!\!
               \right]\!
               :\,
               b,c,d \in \mathbb{R}
               \right\}
      \end{array}
   \right.
   \;\;\;\;  .
\end{equation}
\smallskip

\noindent Under the truth-value gaps (\ref{12}), this necessitates absolutely no truth value for the proposition $\mathrm{D}_1$ in the initial state $|\Psi_{\neg\mathrm{O}}\rangle$:\smallskip

\begin{equation} \label{31} %{Eq.31}
   |\Psi_{\neg\mathrm{O}}\rangle
   =
   \frac{1}{\sqrt{3}}
   \!
   \left[
      \!\!
      \begin{array}{c}
                0 \\
                1 \\
                1 \\
                1
      \end{array}
      \!\!
   \right]
   \!
   \notin
   \left\{
      \begin{array}{l}
         \mathrm{ran}(\hat{P}_{\mathrm{D}_1})\\
         \mathrm{ran}(\neg \hat{P}_{\mathrm{D}_1})
      \end{array}
   \right.
   \iff
   \;
   \{v(\hat{P}_{\mathrm{D}_1})\}
   =
   \{{[\![ \,\mathrm{D}_1\, ]\!]}_v\}
   =
   \varnothing
   \;\;\;\;  .
\end{equation}
\smallskip

\noindent This in turn implies that, in accordance with the probability postulation (\ref{5}), the a priori probability of the simultaneous clicking of the detectors $\mathrm{D}_1^{\mathrm{B}}$ and $\mathrm{D}_1^{\mathrm{A}}$ \textit{must not be equal to zero}:\smallskip

\begin{equation} \label{32} %{Eq.32}
   |\Psi_{\neg\mathrm{O}}\rangle
   \in
   \mathrm{ran}(\neg \hat{P}_{\mathrm{O}})
   \iff
   \left.
      \begin{array}{l}
         {[\![ \,\mathrm{O}\,  ]\!]}_v
         =
         0\\
         \{{[\![ \,\mathrm{D}_1\, ]\!]}_v\}
         =
         \varnothing
       \end{array}
       \!\!
   \right\}
   \implies
   \mathbb{P}\left[ {[\![ \,\mathrm{D}_1\,  ]\!]}_v = 1 \right]
   \neq
   0
   \;\;\;\;  .
\end{equation}
\smallskip

\noindent Contrariwise, when the detectors $\mathrm{D}_1^{\mathrm{B}}$ and $\mathrm{D}_1^{\mathrm{A}}$ click simultaneously and thus the system is found in the state $|{\Psi}_{\mathrm{D}_1}\rangle$, namely,\smallskip

\begin{equation} \label{33} %{Eq.33}
   |{\Psi}_{\mathrm{D}_1}\rangle
   =
   |{\mathrm{D}_1^{\mathrm{A}}}\rangle \otimes |\mathrm{D}_1^{\mathrm{B}}\rangle
   =
   \frac{1}{2}
   \!
   \left[
      \!\!
      \begin{array}{r}
                 1 \\
                -1 \\
                -1 \\
                 1
      \end{array}
      \!\!
   \right]
   \!\!
   \in
   \mathrm{ran}(\hat{P}_{\mathrm{D}_1})
   \implies
   v(\hat{P}_{\mathrm{D}_1})
   =
   {[\![ \,\mathrm{D}_1\,  ]\!]}_v
   =
   1
   \;\;\;\;  ,
\end{equation}
\smallskip

\noindent the proposition asserting that both particles have passed the overlapping arms in their respective interferometers would possess no value at all:\smallskip

\begin{equation} \label{34} %{Eq.34}
   |{\Psi}_{\mathrm{D}_1}\rangle
   \notin
   \left\{
      \begin{array}{l}
         \mathrm{ran}(\hat{P}_{\mathrm{O}})\\
         \mathrm{ran}(\neg \hat{P}_{\mathrm{O}})
      \end{array}
   \right.
   \iff
   \;
   \{v(\hat{P}_{\mathrm{O}})\}
   =
   \{{[\![ \,\mathrm{O}\, ]\!]}_v\}
   =
   \varnothing
   \;\;\;\;  .
\end{equation}
\smallskip

\noindent Accordingly, in supervaluationist (i.e., gappy and yet bivalent) semantics, the question ``Which way did the particle take?'' has no sense.\\

\section{Resolving Hardy's paradox within gapless semantics}  %{IV}

\noindent As it is mentioned in \cite{Beziau}, for gappy semantics, one can construct a gapless (non-classical) semantics in which different degrees of truth would fill out the truth-value gaps.\\

\subsection{Many-valued semantics}  %{IV.1}

\noindent To that end, instead of the truth-value gaps (\ref{12}) consider the gapless valuations, namely,\smallskip

\begin{equation} \label{35} %{Eq.35}
   |\Psi\rangle
   \notin
   \left\{
      \begin{array}{l}
         \mathrm{ran}(\hat{P}_{\diamond})\\
         \mathrm{ran}(\neg \hat{P}_{\diamond})
      \end{array}
   \right.
   \iff
   \;
   v_{\mathbb{P}}(\hat{P}_{\diamond})
   =
   \langle\Psi|\hat{P}_{\diamond}|\Psi\rangle
   \;\;\;\;  ,
\end{equation}
\smallskip

\noindent where the function $v_{\mathbb{P}}$ assigning the truth value to the projection operator $\hat{P}_{\diamond}$ in the state $|\Psi\rangle$ is determined by the probability $\mathbb{P}\left[ {[\![ \,\diamond ]\!]}_v = 1 \right] = \langle\Psi|\hat{P}_{\diamond}|\Psi\rangle$. According to \cite{Pykacz1, Pykacz2}, the value $v_{\mathbb{P}}$ represents the degree to which the proposition $\diamond$ is true before its verification. As $\langle\Psi|\hat{P}_{\diamond}|\Psi\rangle \in \{x \in \mathbb{R} \,:\, 0<x<1 \}$, a semantics defined by the set of the bivaluations (\ref{11}) and the valuations (\ref{35}) is \textit{infinite-valued}. Therewithal, this semantics respects functionality of the evaluation relation; what is more, in this semantics the evaluation relation is a total relation.\\

\noindent Within the said semantics, the truth value of the proposition $\mathrm{D}_1$ in the initial state $|\Psi_{\neg\mathrm{O}}\rangle$ and the truth value of the proposition $\mathrm{O}$ in the final state $|{\Psi}_{\mathrm{D}_1}\rangle$ are given by\smallskip

\begin{equation} \label{36} %{Eq.36}
   |\Psi_{\neg\mathrm{O}}\rangle
   =
   \frac{1}{\sqrt{3}}
   \!
   \left[
      \!\!
      \begin{array}{c}
                0 \\
                1 \\
                1 \\
                1
      \end{array}
      \!\!
   \right]
   \!
   \notin
   \left\{
      \begin{array}{l}
         \mathrm{ran}(\hat{P}_{\mathrm{D}_1})\\
         \mathrm{ran}(\neg \hat{P}_{\mathrm{D}_1})
      \end{array}
   \right.
   \implies
   \;
   v_{\mathbb{P}}(\hat{P}_{\mathrm{D}_1})
   =
   \langle\Psi_{\neg\mathrm{O}}|\hat{P}_{\mathrm{D}_1}|\Psi_{\neg\mathrm{O}}\rangle
   =
   \frac{1}{12}
   \;\;\;\;  ,
\end{equation}

\begin{equation} \label{37} %{Eq.37}
   |{\Psi}_{\mathrm{D}_1}\rangle
   =
   \frac{1}{2}
   \!
   \left[
      \!\!
      \begin{array}{r}
                 1 \\
                -1 \\
                -1 \\
                 1
      \end{array}
      \!\!
   \right]
   \!
   \notin
   \left\{
      \begin{array}{l}
         \mathrm{ran}(\hat{P}_{\mathrm{O}})\\
         \mathrm{ran}(\neg \hat{P}_{\mathrm{O}})
      \end{array}
   \right.
   \implies
   \;
   v_{\mathbb{P}}(\hat{P}_{\mathrm{O}})
   =
   \langle{\Psi}_{\mathrm{D}_1}|\hat{P}_{\mathrm{O}}|{\Psi}_{\mathrm{D}_1}\rangle
   =
   \frac{1}{4}
   \;\;\;\;  .
\end{equation}
\smallskip

\noindent Both of these truth values differ from the classical truth values ``false'' and ``true'', i.e., 0 and 1.\\

\noindent Therefore, in the infinite-valued (i.e., gapless but non-bivalent) semantics the proposition of the simultaneous clicking of the detectors $\mathrm{D}_1^{\mathrm{B}}$ and $\mathrm{D}_1^{\mathrm{A}}$ in the initial state $|\Psi_{\neg\mathrm{O}}\rangle$ is \textit{neither true nor false}. Along the same lines, the proposition affirming that the particles have taken the overlapping ways $\mathrm{O}^{\mathrm{B}}$ and $\mathrm{O}^{\mathrm{A}}$ is neither true nor false in the final state $|{\Psi}_{\mathrm{D}_1}\rangle$.\\

\subsection{Weak-valued semantics}  %{IV.2}

\noindent Another possibility to construct a gapless semantics presents weak values of the projection operators. To be sure, consider the ``weak'' valuations\smallskip

\begin{equation} \label{38} %{Eq.38}
   |\Psi\rangle
   \notin
   \left\{
      \begin{array}{l}
         \mathrm{ran}(\hat{P}_{\diamond})\\
         \mathrm{ran}(\neg \hat{P}_{\diamond})
      \end{array}
   \right.
   \iff
   \;
   v_{w}(\hat{P}_{\diamond})
   =
   \frac{
      \langle\Phi|\hat{P}_{\diamond}|\Psi\rangle
   }
   {
      \langle\Phi|\Psi\rangle
   }
   \;\;\;\;  ,
\end{equation}
\smallskip

\noindent where the bra $\langle\Phi|$ is some vector of the dual space ${\mathcal{H}}^{*}$ associated with the given quantum system.\\

\noindent Had the quantum system been prepared in the state $|\Psi\rangle$ lying in the column or null space of the projection operator $\hat{P}_{\diamond}$, such valuations would have coincided with the bivaluations (\ref{11}). What is more, had the bra $\langle\Phi|$ been the Hermitian conjugate of the ket $|\Psi\rangle$, i.e., $\langle\Phi| = \langle\Psi|$, the value $v_{w}(\hat{P}_{\diamond})$ would have lain in the range $\{x \in \mathbb{R} \,:\, 0<x<1 \}$.\\

\noindent Given that neither of these statements is fulfilled, one can deduce that a semantics defined as the set of the bivaluations (\ref{11}) and the weak valuations (\ref{38}) is not bivalent and, unlike many-valued semantics, may include ``truth degrees'' lying beyond the range $[0,1]$. This last detail makes any conventional interpretation of truth degrees inapplicable to the weak truth values.\\

\noindent Even so, an interpretation of the weak truth degrees, uncontroversial in a sense, can be based on the observation made in \cite{Vaidman} and the comments in \cite{Svensson}: According to them, non-zeroness (respectively, zeroness) of the value $v_{w}(\hat{P}_{\diamond})$ can be interpreted as the divergence (conformity) of the proposition $\diamond$ from (to) \textit{the false}, i.e., a trivial proposition whose truth value being always false.\\

\noindent Take, for example, the proposition $\mathrm{D}_1$ in the initial state $|\Psi_{\neg\mathrm{O}}\rangle$ and the proposition $\mathrm{O}$ in the final state $|{\Psi}_{\mathrm{D}_1}\rangle$: Under the weak valuations (\ref{38}), they can be evaluated as\smallskip

\begin{equation} \label{39} %{Eq.39}
   |\Psi_{\neg\mathrm{O}}\rangle
   \!
   \notin
   \left\{
      \begin{array}{l}
         \mathrm{ran}(\hat{P}_{\mathrm{D}_1})\\
         \mathrm{ran}(\neg \hat{P}_{\mathrm{D}_1})
      \end{array}
   \right.
   \implies
   \;
   v_{w}(\hat{P}_{\mathrm{D}_1})
   =
   \frac{
      \langle{\Psi}_{\mathrm{D}_1}|\hat{P}_{\mathrm{D}_1}|\Psi_{\neg\mathrm{O}}\rangle
   }
   {
      \langle{\Psi}_{\mathrm{D}_1}|\Psi_{\neg\mathrm{O}}\rangle
   }
   =
   1
   \;\;\;\;  ,
\end{equation}

\begin{equation} \label{40} %{Eq.40}
   |{\Psi}_{\mathrm{D}_1}\rangle
   \!
   \notin
   \left\{
      \begin{array}{l}
         \mathrm{ran}(\hat{P}_{\mathrm{O}})\\
         \mathrm{ran}(\neg \hat{P}_{\mathrm{O}})
      \end{array}
   \right.
   \implies
   \;
   v_{w}(\hat{P}_{\mathrm{O}})
   =
   \frac{
      \langle\Psi_{\neg\mathrm{O}}|\hat{P}_{\mathrm{O}}|{\Psi}_{\mathrm{D}_1}\rangle
   }
   {
      \langle\Psi_{\neg\mathrm{O}}|{\Psi}_{\mathrm{D}_1}\rangle
   }
   =
   0
   \;\;\;\;  .
\end{equation}
\smallskip

\noindent In line with the said interpretation, these weak truth degrees mean that despite the falsity of the proposition $\mathrm{O}$ in the initial state $|\Psi_{\neg\mathrm{O}}\rangle$, the proposition of the simultaneous clicking $\mathrm{D}_1$ cannot be regarded as false in this state. By the same token, the fact that in the final state $|{\Psi}_{\mathrm{D}_1}\rangle$ the proposition $\mathrm{D}_1$ diverges from the false does not imply that in the same state the proposition $\mathrm{O}$ differs from the false as well.\\

\noindent That is to say, in the weak-valued (i.e., gapless and non-bivalent) semantics, it holds that\smallskip

\begin{equation} \label{41} %{Eq.41}
   {[\![ \,\mathrm{O}\,  ]\!]}_{v}
   =
   0
   \;
   \nRightarrow
   \;
   {[\![ \,\mathrm{D}_1\,  ]\!]}_{v_{w}}
   =
   0
   \;\;\;\;  ,
\end{equation}   

\begin{equation} \label{42} %{Eq.42}
   {[\![ \,\mathrm{D}_1\,  ]\!]}_{v}
   \neq
   0
   \;
   \nRightarrow
   \;
   {[\![ \,\mathrm{O}\,  ]\!]}_{v_{w}}
   \neq
   0
   \;\;\;\;  .
\end{equation}

\vspace*{7mm}

\noindent As it follows, any of the mentioned in this paper non-classical semantics resolve Hardy's paradox.\\

\section*{Acknowledgment}
\noindent The author would like to express acknowledgment to the anonymous referee for the constructive remarks contributing to the improvement of this paper.\\

\end{document}